\documentclass[aps,prl,preprint]{revtex4}
\usepackage{amssymb}
\usepackage{graphicx}
\def\beqa{\begin{eqnarray}}
\def\eeqa{\end{eqnarray}}
\def\be{\begin{equation}}
\def\ee{\end{equation}}

\begin{document}

\title{A matter--dominated cosmological model with variable $G$ and $\Lambda$ and its
confrontation with observational data}

\author{Ester Piedipalumbo}
\email[E-mail at: ]{ester@na.infn.it}
\affiliation{Dipartimento di Scienze Fisiche, Universit\`{a} Federico II\\
and \\ Istituto Nazionale di Fisica Nucleare, Sez. di Napoli,\\
Complesso Universitario di Monte S. Angelo,\\ Via Cintia, Ed. G, I-80126 Napoli, Italy}

\author{Giampiero Esposito}
\affiliation{Istituto Nazionale di Fisica Nucleare, Sez. di Napoli,\\
Complesso Universitario di Monte S. Angelo,\\ Via Cintia, Ed. G, I-80126 Napoli, Italy}

\author{Claudio Rubano}
\affiliation{Istituto Nazionale di Fisica Nucleare, Sez. di Napoli,\\
Complesso Universitario di Monte S. Angelo,\\ Via Cintia, Ed. G, I-80126 Napoli, Italy}

\author{Paolo Scudellaro}
\affiliation{Dipartimento di Scienze Fisiche, Universit\`{a} Federico II\\
and \\ Istituto Nazionale di Fisica Nucleare, Sez. di Napoli,\\
Complesso Universitario di Monte S. Angelo,\\ Via Cintia, Ed. G, I-80126 Napoli, Italy}

\date{\today}

\begin{abstract}
In the framework of renormalization-group improved cosmologies, we analyze both theoretically and
observationally the exact and general solution of the matter--dominated cosmological equations, using the
expression of $\Lambda = \Lambda (G)$ already determined by the integration method employed in a previous paper.
A rough comparison between such a model and the concordance $\Lambda$CDM model as to the magnitude--redshift
relationship has been already done, without showing any appreciable differences. We here perform a more refined
study of how astrophysical data (Union2 set) on type-I supernovae, gamma ray bursts (in a sample
\textit{calibrated} in a model independent way with the SneIa dataset), and gas fraction in galaxy clusters
(using a sample of Chandra measurements of the X-ray gas mass fraction) affect the model and constrain its
parameters. We also apply a cosmographic approach to our cosmological model and estimate the cosmographic
parameters by fitting both the supernovae and the gamma ray bursts datasets. We show that this matter-dominated
cosmological model with variable Newton parameter and variable cosmological term is indeed compatible with the
observations above (on type Ia supernovae, the gamma ray bursts Hubble diagram, and the gas mass fraction in
X-ray luminous galaxy clusters). The cosmographic approach adopted confirms such conclusions. Finally, it seems
possible to include radiation into the model, since numerical integration of the equations derived by the
presence of both radiation and matter shows that, after inflation, the total density parameter is initially
dominated by the radiation contribution and later by the matter one.
\end{abstract}


\maketitle

\section{Introduction}

Pushed on by the overwhelming flow of observational data in the last fifteen years, most cosmologists today
agree on a well defined cosmological paradigm, based on General Relativity plus a cosmological constant
$\Lambda$. This paradigm is known as the Concordance Cosmological Model \cite{d07} and accounts not only for the
early formation of large--scale structures but also for the more recently discovered stage of acceleration of
the universe. As to the building up of galaxies and galaxy clusters, it has been necessary to introduce an
ingredient like dark matter, which was first employed to succeed in describing rotational curves in spiral
galaxies \cite{dm} (for an alternative view, see for example the work in Ref. \cite{Lusa10}). On the other hand,
cosmic acceleration requires the consideration of the so--called dark energy as the major ingredient of the
cosmic content \cite{de}, $\Lambda$ being just the simplest way to consider it. Both of them sum up to more than
$95 \%$ of the matter--energy constituents around us.

Such dark energies, on the other hand, have simply hidden the fundamental issues, since so far no exhaustive
physical explanation for them has been put on firm theoretical and experimental ground. This has led to many
alternative ways to reproduce the astrophysical phenomena cited above, not only by introducing theoretically
well motivated new particles and fields, but also by means, for instance, of possible geometrical changes of the
spacetime structure ($f(R)$ theories are well known examples of this kind of proposals \cite{sal} \cite{far}.

Without entering any details of so many attempts to describe the cosmological behaviour, we shall here consider
only some aspects of one of them, i.e. that stemming from the possibility that not only the cosmological term
$\Lambda$ could vary with space and time, but also the gravitational coupling $G$. In this context, one way to
achieve the physical realization of such assumptions is to study cosmological dynamics by analyzing
``renormalization group (RG) induced'' quantum effects, which drive the dimensionless cosmological ``constant''
$\lambda(k)$ and Newton ``constant'' $g(k)$ from an ultraviolet attractive fixed point \cite{Reut94}
\cite{Reut98} \cite{Soum99} \cite{Wett01} \cite{Berg02} \cite{Laus02a}. In the exact theory, such a non-Gaussian
ultraviolet fixed point implies its nonperturbative renormalizability \cite{Laus02a} \cite{Reut02b}
\cite{Laus02b} \cite{Bab05} \cite{Cod06} \cite{Bona05} \cite{Nied03} \cite{Nied02} \cite{Forg02}. As a result,
this Renormalization Group --improved framework describes gravity at a typical distance scale $\ell\equiv
k^{-1}$, and introduces an effective average action $\Gamma_k[g_{\mu\nu}]$ for Euclidean quantum gravity
\cite{Reut98}, finally implying an exact functional Renormalization Group equation for the $k$--dependence of
$\Gamma_k$. This framework is usually known as \emph{quantum Einstein gravity}. Within it, one can get an
explicit $k$--dependence of both the running Newton and cosmological terms $G(k)$ and $\Lambda(k)$, which can be
relevant both for the initial Planck era and the structure of black hole singularities \cite{Bona02a}
\cite{Bona99} \cite{Bona00}.

Taking into account its inherent infrared divergences, quantum Einstein gravity can be subject to strong
renormalization effects even at very large distances. In cosmology, such effects lead to a dynamical relaxation
of $\Lambda$ and can also be assumed to deal with the cosmological constant problem \cite{Tsam93}. Viewing the
late accelerated expansion of the universe as a renormalization group evolution near a non--Gaussian infrared
fixed point \cite{Bona02b} (although the actual existence of an infrared fixed point has not been proved as
yet), one can assume that the transition between standard FLRW cosmology and accelerated Renormalization Group
driven expansion occurs at the time when the fixed point is almost reached. As to this, some agreement has been
found between this kind of model and early SNeIa observations \cite{Bent04}.

As a matter of fact, for a homogeneous and isotropic universe, it is possible to identify $k$ with the inverse
of cosmological time, $k \propto 1/t$ \cite{Bona02a} \cite{Bona02b}, hence deriving a dynamical evolution for
$G(k)$ and $\Lambda(k)$ induced by their Renormalization Group  running. The Arnowitt--Deser--Misner (ADM)
formulation \cite{Bona04} builds a modified action functional which reduces to the Einstein--Hilbert action when
$G$ is constant. Within such a framework, and always assuming homogeneity and isotropy, one can obtain a
power--law growth of the scale factor for pure gravity and for a massless $\varphi^4$ theory, in agreement with
what is known on fixed--point cosmology. On the other hand, by means of the so-called Noether Symmetry Approach
\cite{deritis90} \cite{cap96}, in Ref. \cite{Bona07a} we have also proposed solutions for the pure gravity case,
which mimic inflation without introducing a scalar field in the cosmic content. In Ref. \cite{Bona08}, for
gravity with a scalar field, this approach only succeeds in fixing the expressions of $\Lambda(G)$ and
$V(\varphi)$ as $\Lambda \propto G$ and $V \propto \varphi^2$, respectively, while the transformed cosmological
equations derived by means of the method do not seem to be easily solvable \cite{Bona08}, therefore giving no
new insight into possible solutions.

In what follows, we take again into account the exact solutions of the flat dust matter-dominated cosmological
equations (without any scalar field), already investigated in Ref. \cite{Bona07b} by means of the Noether
Symmetry Approach, and by using an expression for $\Lambda=\Lambda(G)$ determined by the method itself. After
briefly reviewing the theoretical model, we show that our cosmological model is compatible with various recent
observational data, in particular with the observations of type Ia supernovae (SNeIa) (we use the recently
updated SNeIa sample, referred to as Union2 \cite{amanullah}, containing 557 SNeIa spanning the redshift range
$0.015 \le z \le 1.55$.), the Gamma Ray Bursts Hubble diagram (GRBs HD) (we use a sample \textit{calibrated} in
a model independent way with the SneIa dataset \cite{MEC10}), and the gas mass fraction in X-ray luminous galaxy
clusters (we use a sample of Chandra measurements of the X-ray gas mass fraction in 42 hot ($kT>5$keV), X-ray
luminous, dynamically relaxed galaxy clusters spanning the redshift range $0.05< z<1.1$ \cite{allen08}).

We are aware that such observational tests cannot definitively distinguish our model from the standard
$\Lambda$CDM one, at least because of the presence of still large observational errors. However, it provides  a
crucial and necessary test of reliability.

Looking at future studies on this, we finally apply to our cosmological model a cosmographic approach, which can
indeed contribute to select realistic models without imposing arbitrary choices {\it a priori}. As a matter of
fact, cosmography and its reliability are based on the assumptions that the universe is spatially homogeneous
and isotropic on large scale, and luminosity distance can be ``tracked'' by the derivative series of the scale
factor $a(t)$. We actually estimate the cosmographic parameters here derived by fitting both the SNeIa Union2
dataset and the \textit{calibrated} GRBs HD.

We also begin to study how our model is affected by the inclusion of radiation into the cosmic content. It in
fact turns out that, by performing the numerical integration of the equations so rewritten, the total density
parameter is initially dominated by the radiation contribution and later on by the matter one, leaving then
space to the now observed accelerated stage.

The scheme of the paper is as follows. In Section 2 we summarize the Lagrangian formulation used to derive the
Renormalization--Group improved Einstein cosmological equations with the ordinary matter energy--momentum
tensor, as well as the results deduced from the Noether symmetry found in Ref. \cite{Bona07b}. In Sections 3, 4
and 5 we present the comparison of theoretical predictions with observational data. Section 6 is then devoted to
the cosmographic approach, and Section 7 to inclusion of radiation into the model. Finally, some conclusions are
drawn in Section 8.

\section{Theoretical model}

Let us consider the approach outlined in Ref. \cite{Bona04} \cite{Bona07a} \cite{Bona07b} \cite{Bona08} and
there applied to models of gravity with variable $G$ and $\Lambda$ in the context of quantum Einstein gravity.
It is known that, in a homogeneous and isotropic universe, an independent dynamical $G$ is equivalent to
metric-scalar gravity already at classical level \cite{Cap97} \cite{Cap98}, while independent variations (with
position and time) of $G$ and $\Lambda$ can lead to pathological situations. Indeed, if $\Lambda$ were an
independent variable, one should write that the momentum conjugate to it vanishes, and the preservation in time
of this primary constraint would imply a vanishing lapse function and hence a ``collapse'' of spacetime geometry
\cite{Bona04}. All this in fact leads to assume a generic functional dependence $\Lambda = \Lambda(G)$
\cite{Bona04}.

In the matter--dominated case in a flat homogeneous and isotropic cosmology (with a signature $-,+,+,+$ for the
metric, lapse function $N=1$ and shift vector $N^{i}=0$), as in Ref. \cite{Bona07b}, we start from the
Lagrangian
\begin{equation} \label{2}
L = \frac{1}{8 \pi G} \left( -3a\dot{a}^2 - a^3 \Lambda + \frac{1}{2}\mu a^3 \frac{\dot{G}^2}{G^2} \right) -
Da^{-3(\gamma-1)}\,,
\end{equation}
where $G=G(t)$ and $\Lambda=\Lambda(G(t))$, while dots indicate time derivatives, and $\mu$ is a nonvanishing
interaction parameter introduced in Ref. \cite{Bona04} and also used in Ref. \cite{Bona07b}, where it was shown
that $\mu > 2$. The matter contribution is of course given by $L_m \equiv - Da^{-3(\gamma-1)}$, with $1 \leq
\gamma \leq 2$; here, we have to take $\gamma = 1$ for dust, while $D$ is a suitable integration constant
connected to the matter content. As shown in Ref. \cite{Bona07b}, from Eq. (\ref{2}) we get the Euler--Lagrange
equations for $a$ and $G$
\begin{equation} \label{3}
\frac{\ddot{a}}{a} + \frac{\dot{a}^2}{2a^2} - \frac{\Lambda}{2} - \frac{\dot{a}\dot{G}}{a G} + \frac{\mu
\dot{G}^2}{4G^2} = 0\,,
\end{equation}
\begin{equation} \label{4}
\mu \ddot{G} - \frac{3}{2}\mu \frac{{\dot{G}}^2}{G} + 3\mu \frac{\dot{a}}{a}\dot{G} +
\frac{G}{2} \left( -6\frac{\dot{a}^2}{a^2} -2\Lambda + 2G\frac{d\Lambda}{d G} \right) = 0\,.
\end{equation}
The Hamiltonian constraint \cite{Bona07b}
\begin{equation} \label{5}
\frac{\dot{a}^2}{a^2} - \frac{\Lambda}{3} - \frac{\mu}{6}\frac{\dot{G}^2}{G^2} -\frac{8\pi G}{3}Da^{-3}= 0
\end{equation}
is equivalent to the constraint on the \emph{energy} function associated with $L$ \cite{deritis90}\cite{cap96}
\cite{Bona07a} \cite{Bona07b} \cite{Bona08}
\begin{equation}
\label{5bis} E_L \equiv \frac{\partial L}{\partial \dot{a}}\dot{a} + \frac{\partial L}{\partial \dot{G}}\dot{G} - L = 0\,.
\end{equation}

For matter, the dust case involves a zero pressure, $p_m=0$, and an energy density $\rho_m = Da^{-3}$, so that
the matter term in the Lagrangian is simply a constant. It therefore has no effect on the equations of motion
with respect to the pure gravity case; nevertheless, it has to be considered, since it occurs in the constraint
equation (\ref{5bis}). The system of equations of motion can then be solved \cite{Bona07b} by using the Noether
Symmetry Approach \cite{deritis90} \cite{cap96}, in which we consider $L$ as a point Lagrangian, a function of
the variables $a$ and $G$, and their first derivatives \cite{deritis90} \cite{cap96} \cite{Bona07a}
\cite{Bona07b}. We have already shown that a consistent choice of the function $\Lambda = \Lambda(G)$ leads to
the existence of a Noether symmetry for the Lagrangian \cite{Bona07b}. As a matter of fact, in the
matter--dominated case we get the same Noether symmetry as in the pure gravity situation, so that, by using the
same transformations introduced in this latter case, one can write $a=a(t)$ and $G=G(t)$ as therein, now just
updating the energy constraint. (For more details, see Ref. \cite{Bona07b}.)

The dynamics of $\Lambda$ is coupled with that of $G$ and is driven by the equation
\begin{equation} \label{lameq}
2 (1-J)\Lambda + G\frac{d\Lambda}{d G}=0,
\end{equation}
where the parameter $J$ is an arbitrary constant related to the interaction factor $\mu$ by the relation $\mu
=\frac{2}{3}(3-2 J)^2\neq 0, \frac{2}{3}$. This equation admits the solution
\begin{equation} \label{3.34}
\Lambda = \Lambda (t; n) = W G^{\frac{1}{1-3n}}\,,
\end{equation}
$W$ being an integration constant and $n(J) \equiv \frac{3-2J}{6(1-J)}$. It turns out that $W$ is related to the
present value of $G$ \cite{Bona07b}, and we might determine $W$ in order to get $G_0 \equiv G_N \equiv G_{\rm
Newton}$.

Furthermore, we fix time scale and origin so as to get $a(0)=0$. Thus, we find \cite{Bona07b}
\begin{eqnarray}
a = a(t) & = & A \left(t^{\frac{1}{6 n-1}+1}
\left(B+t^{\frac{1}{6 n-1}+1}\right)\right)^n, \\
G = G(t) & = & C \left( t^2 + B t^{\frac{2-6n}{1-6n}} \right)^{3n-1}\,,
\end{eqnarray}
where we define the constants
\begin{eqnarray}
A & \equiv & A(n, W) \equiv
12^{\frac{n(1+6n)}{1-6n}}n^{\frac{12n^2}{1-6n}}
(6n-1)^{\frac{12n^2}{6n-1}}(12n-1)^{-n}{W}^n\label{eqA} \,,\\
B & \equiv & B(n, W, D) \equiv {W}^{-1} \left[ 2^{\frac{3(1-10n)}{1-6n}}(3n)^{\frac{6n}{6n-1}}
(6n-1)^{\frac{1-12n}{6n-1}}(12n-1)\pi D \right]\label{eqB}\,,\\
C & \equiv & C(n, W) \equiv (6n-1)^{2(3n-1)}\left[ 12n^2(12n-1) \right]^{1-3n} {W}^{3n-1} \,.
\end{eqnarray}
Here we point out that the asymptotic time behavior of the scale factor is characterized by the two exponents
\begin{equation}
p_{1} \equiv \frac{12n^2}{(6n-1)}\,, \; p_{2}  \equiv \frac{6n^2}{(6n-1)}\,, \label{p2}
\end{equation}
and we want that $p_1 >1$ and $p_2 < 1$ (in order to obtain early matter domination and a later accelerated
evolution ), which implies that we find a limited range of variability for the $n$ parameter\footnote{It is
worth noting that such a constraint on the parameters $p_i$ of our model is not strictly necessary, but it has
been used, already in Ref. \cite{Bona08}, as an instrument to fix somehow the reachable regions in the space of
parameters (at least for what it concerns $n$). This means that one could perform our analysis leaving the $p_i$
parameters, and then $n$, completely free}, $(3-\sqrt{3})/6 < n< (3+\sqrt{3})/6$.  We then see that, when $D=0$,
one has $B=0$, and we recover the same results obtained in Ref. \cite{Bona08} for the pure gravity model.

Eqs. (\ref{eqA}) and (\ref{eqB}) make it possible to obtain $D$ and $W$ as functions of $A$, $B$ and $n$, i.e.
\begin{eqnarray}
D &\equiv& \frac{2^{\frac{2}{6 n-1}-3} 3^{\frac{1}{6 n-1}} n^{\frac{1}{6 n-1}+1} (6
   n-1)^{\frac{1}{1-6 n}} A^{\frac{1}{n}}B}{\pi }\,,\label{eqD}\\
   W &\equiv &  12^{\frac{2}{6 n-1}+1} n^{\frac{12 n}{6 n-1}} (6 n-1)^{\frac{12 n}{1-6 n}} (12 n-1)
   A^{\frac{1}{n}}\,.
\end{eqnarray}
To make things analytically simpler and obtain a better control of the space of parameters related to the
integration constants, we set the present time $t_0 = 1$. This fixes the scale of time according to the
(unknown) age of the universe. In other words, this means that we are using the age of the universe, $t_0$, as a
unit of time, and the whole history of the universe has been {\it squeezed} to the range of time  $[0,1]$. We
then set $a_0 = a(1) = 1$, which is standard, and  finally $ H_0= H(1)\simeq 1$. Because of our choice of time
unit, it turns  out that {\it our $H_0$ is not the same as the Hubble constant} which appears in the standard
FLRW model.

Such choices introduce a constraint between $A$ and $B$:
\begin{equation}
A = (1 + B)^{-n}\,.
\end{equation}
On choosing to  normalize $a_0$, the definition of the redshift $z \equiv a_0/a - 1=  1/a - 1$ yields
\begin{equation}\label{zt}
z = z(t) = (B+1)^n t^{\frac{6 n^2}{1-6 n}} \left(B+t^{\frac{1} {6 n-1}+1}\right)^{-n}-1\,,
\end{equation}
depending only on the parameter $n$. It turns out that we obtain the following expression for $H_0$ and $\Lambda_0$:

\begin{equation} \label{h0eq}
H_0\equiv H(t_0 = 1) =\frac{6 (2 + B) n^2 }{(1 + B) (-1 + 6 n)}\,,
\end{equation}
\begin{equation}
\Lambda_0\equiv \Lambda(t_0 = 1)= -\frac{6 (B+2) n^2 (B+1)^{n-1}}{6 n-1}\,.
\end{equation}
We have constrained the parameters of our model in order to have  $G_{0}=G_N=1$.
Therefore, the density parameter of matter is
\begin{equation} \label{omegam0}
\Omega_{m} \equiv \frac{8\pi G(t,B,n)D}{3{H(t,B,n)}^2}\left|_{t=1}\right.\equiv \frac{8\pi G_0
D}{3{H_0}^2}= \frac{B (B+1) (6 n-1)}{9 (B+2)^2 n^2}\,,
\end{equation}
and it turns out that the constraint
\begin{equation}
\Omega_{m}+\Omega_{\Lambda 0}+\Omega_{G 0} = 1
\end{equation}
is satisfied. Moreover, as already pointed out in Ref. \cite{Bona06}, we do not expect that $G_0 \equiv G_N
\equiv G_{\rm Newton}$, even if we fixed the parameters in order to get it. Anyway, we have to consider that
small differences between $G_0$ and $G_N$ could imply $G_0$ varying a lot in time, with relevant effects on the
evolution of the universe. It is worth noting that $G$ has a special role in the subject of time variation of
the fundamental parameters. Actually a  dependence of $G$ on time may point out to violation of the strong
equivalence principle, but not necessarily of the Einstein equivalence principle, whereas the nonconstancy of
the other ``constants``, like the electroweak or strong coupling constants, necessarily represents a violation of
the equivalence principle in both its forms. Here we will only discuss some aspects of the response of
primordial abundances to the time variations of $G$, as expected in our model, on the basis of previous analysis
on the subject performed in Ref. \cite{Bambi05}.

\begin{figure}
\includegraphics[width=8 cm, height=6 cm]{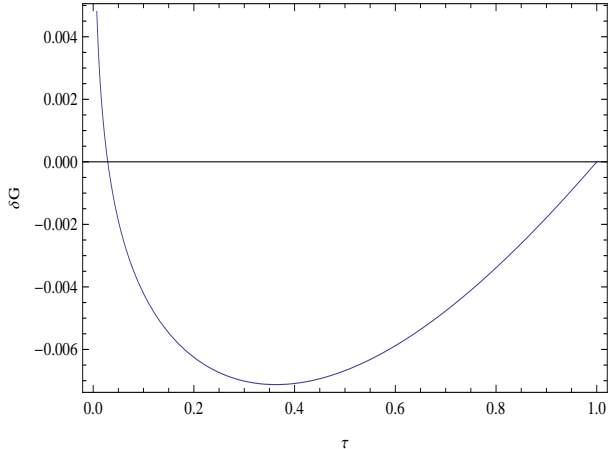}
\caption{Time evolution for the relative variation $\delta G$ for our model, with $B=2.77$, $n=0.32$. As we will
discuss in the next sections such values are fully consistent with the constraints resulting from other
cosmological datasets.}\label{GvariableBBN}
\end{figure}

It turns out that the production of each element responds in its own way to a variation $\delta G$ of the Newton
constant. A general study, that can account for a time dependence of $G$ during the BBN period, requires the
introduction of suitable functions which describe the response of each element to an arbitrary time--dependent
modification of the early universe expansion rate, and it is out of the aim of the present work. So, we simply
discuss the observational bounds on the possible variations of the gravitational constant in the early universe,
considering the best limit (at $3 \sigma$), $\delta G = 0.09 ^{+0.22}_{-0.19}$, obtained in Ref. \cite{Bambi05},
by combining $^{2}H$ observational results with the measurements of the baryon to photon ratio obtained from CMB
and LSS data\footnote{This limit refers to the value of $G$ when the temperature of the universe is $0.02\leq
T\leq 0.2\,$ MeV (i.e. during and immediately after the d-bottleneck epoch) and is consistent with the standard
assumption that $G$ has lightly varied during the evolution of the universe.}.  We will postpone to a
forthcoming paper the detailed analysis of the dependence of the various elemental abundances on the time
variation of $G$ for our model. It turns out that our model can satisfy such a best fit limit, provided that the
$n$ parameter is appropriately selected (the role of $B$ is only marginal with respect to this strong
constraint), and compatible with the other basic cosmological observations, as shown in Fig. \ref{GvariableBBN}.

To further investigate this issue, we first evaluate the fractional time rate of change of $G$ \be
\frac{\dot{G}}{G_0}\sim \times 10^{-2}\,. \ee Here, it is important to remember that we are using the age of the
universe as unit, so that the effective rate is of order $10^{-13}$ yr$^{-1}$, in agreement with the
observations (see for instance Ref. \cite{pitjeva05}).

We can thus say that, eventually, the running of both the gravitational coupling $G = G(t)$ and the cosmological
term $\Lambda = \Lambda (t)$ induced by quantum effects appears to yield both a primordial inflation (soon after
the universe exits the region where the attraction basin of the non-Gaussian (ultraviolet) fixed point works
\cite{Bona06}), and  a later inflationary epoch in a \emph{matter--dominated} period of the expansion of the
universe. This appears to be interesting since it is obtained always without having to introduce any scalar
field in the cosmic content \cite{Bona07b}.

\subsection{Re-parametrization of the model}\label{rep}

Let us now exhibit a re-parametrization of our model in terms of $H_0$ and $\Omega_m$ instead of $B$ and $n$. On
inverting the systems of Eqs. (\ref{h0eq}) and (\ref{omegam0}) it is indeed possible to recover the parameters
$B$ and $n$ as functions of $H_0$ and $\Omega_m$; actually, we have that
\begin{eqnarray}
  B &=& \frac{6 H_{0} \Omega_m}{2-3 H_{0} \Omega_m}\,, \label{inv1}\\
  n &=& \frac{1}{24} \left(\sqrt{3} \sqrt{H_0 (3 H_0 \Omega_m + 2)
\left(9 H_0^2 \Omega_m +6 H_0-8 \right)}+9 H_0^2 \Omega_m +6 H_0\right)\,.\label{inv2}
\end{eqnarray}

\begin{figure}
\includegraphics[width=6 cm, height=6  cm]{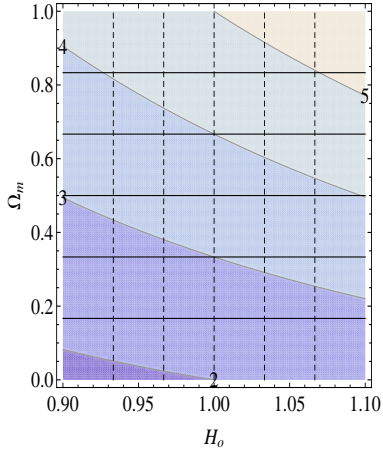}
\caption{Allowed regions (in light blue) of the space of parameters $n$ and $B$.}\label{rHom2}
\end{figure}

The conditions on the two exponents $p_1$ and $p_2$ (that is $p_1 > 1$ and $p_2 < 1$) give rise to a constraint
on the space of parameters for $H_0$ and $\Omega_m $, as shown in Fig. \ref{rHom2}; actually, it turns out
that
\begin{equation}\label{constrain}
2 < H_0 (3 H_0 \Omega_m +2) < 4\,.
\end{equation}
It is worth noting that the space of parameters $H_0$ and $\Omega_m $ is reasonably deducible from physical
arguments. As a matter of fact, $\Omega_m $ varies in the range $\left[0, 1\right]$, and the range of variation
for $H_0$ can be inferred by assuming that the age of the universe can be written in the following way:
\begin{equation}
t_0=\gamma \times  1 Gy=\,3.15\cdot \,10^{16}\,\gamma   \,  s\,,
\end{equation}
where $\gamma$ is a constant to be determined by astronomical observations. With this definition the value of
$H_0$ can be related to the small $ h={\bar H}_0/100 $ of the standard FLRW model. It turns out that
\begin{equation}\label{eq:hubble-conversion}
  H_0= 0.1\,   h\,   \, \gamma\,.
\end{equation}
If we accept that $t_0=\,13.76 \pm 0.11 \, Gy$ and  $h= 0.71 \pm 0.014$, as given by  WMAP7 \cite{jarosik}, then
the region of variability at $2 \sigma$ for $H_0$ turns out to be $\left(0.92\,, 1.03 \right)$. It is
interesting to note that the constraint in Eq. (\ref{constrain}) is satisfied in the whole domain of $H_0$ and
for $ 0.06 <\Omega_m < 0.6$.

\section{Constraints from recent SNeIa observations}

\label{snIa} Over the last years the confidence in type Ia supernovae as standard candles has been steadily
growing.  Actually, it was just the SNeIa observations that gave the first strong indication of an accelerating
expansion of the universe, which can be accounted for by assuming the existence of some kind of dark energy or
nonzero cosmological constant \cite{Schmidt}. Since 1995 two teams of astronomers - the High-Z Supernova Search
Team and the Supernova Cosmology Project - have been discovering type Ia supernovae at high redshifts. First
results of both teams were published in Refs. \cite{Schmidt} and \cite{per+al99}.

As to a first comparison od theory with observations, we here consider the recently updated Supernovae Cosmology
Project \textit{Union2} compilation \cite{amanullah}, which is an update of the original \textit{Union}
compilation, now bringing together data for $719$ SNeIa, drawn from $17$ datasets. Of these, $557$ SNeIa,
spanning the redshift range ($0.015 \le z \le 1.55$.), pass usability cuts and outliers removal, and form the
final sample used to constrain our model. We actually compare the \textit{theoretically\thinspace\ predicted}
distance modulus $\mu(z)$ with the \textit{observed} one, through a Bayesian approach, based on the use, as
merit function, of the likelihood $\mathcal{L}=\exp{\left( -\frac{1} {2}\chi^{2}\right)  }$. The distance
modulus is defined by
\begin{equation}
\mu \equiv m-M = 5\log{d_{L}(z)} +5\log \left({c\over 100 h}\right)+25\,, \label{eq:mMr}
\end{equation}
where $m$ is the appropriately corrected apparent magnitude including reddening, K correction etc., $M$ is the
corresponding absolute magnitude, and $d_{L}$ is the luminosity distance in Mpc. However, in our cosmological
model with variable $G$ and $\Lambda$, it is important also to include in  Eq. (\ref{eq:mMr}) corrections
describing the effect of the time variation of the gravitational constant $G$ on the luminosity of high redshift
supernovae. If the local value of $G$ at the spacetime position of the most distant supernovae differs from
$G_N$, this could in principle induce a change in the Chandrasekhar mass $M_{ch}\propto G^{-{3\over 2}}$. Some
analytical models of the supernovae light curves predict that the peak luminosity is proportional to the mass of
nickel produced during the explosion, which is a fraction of the Chandrasekhar mass. The actual fraction varies
in different scenarios, but the physical mechanism of type Ia supernovae explosion always relates the energy
yield to the Chadrasekhar mass. Assuming that the same mechanism for the ignition and the propagation of the
burning front is valid for SNeIa at high and low redshifts, the predicted apparent magnitude will be fainter by
a quantity \cite{Gat01}
\begin{equation}
\Delta M_{G}={15\over 4}\log\left(G\over G_{0}\right)\label{corgef}.
\end{equation}
Taking this into account the distance modulus becomes
\begin{equation}
m-M=5\log{d_{L}(z)} +5\log\left({c\over 100 h}\right)+25 + \Delta M_{G}\,. \label{eq:modg}
\end{equation}
The presence of this correction allows us to appropriately test our model by using the SNeIa sample \cite{Gat01}
\cite{uz}\cite{mecp}.

\begin{figure}
\includegraphics[width=6 cm, height=6 cm]{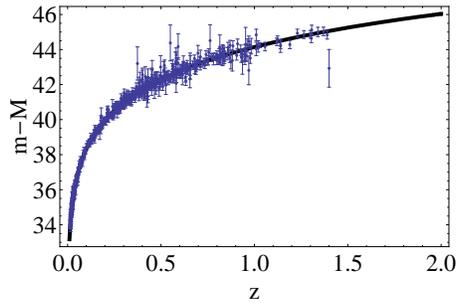}
\caption{The fit of the Union2 data set with the theoretical modulus of distance $\mu$ with respect to redshift
$z$.} \label{f3}
\end{figure}

In our flat and homogeneous cosmological model the luminosity distance can be expressed as an integral of the
Hubble function as follows:
\begin{eqnarray}\label{luminosity}
d_L (z) &=& {c \over H_0}(1+z)\int^{z}_{0}{1\over H(\zeta)}d\zeta\,,
\end{eqnarray}
where $H(z)$ is the Hubble function expressed in terms of redshift.
It turns out that the luminosity distance can be expressed as a function of time in the following way:
\begin{eqnarray}
&& d_L (t)=-\frac{6 (B+2) n^2 (t-1) (B+1)^{2 n-1} \left(t^{\frac{6 n}{6 n-1}} \left(B+t^{\frac{6
   n}{6 n-1}}\right)\right)^{-n}}{6 n-1}\\
   && \times \left( \frac{1}{6 n^2-6 n+1}\left((-1 + 6 n)
\left(\left(\frac{1}{B}+1\right)^n (B+1)^{-n} \, _2F_1\left[-n+1-\frac{1}{6
   n},n;-n+2-\frac{1}{6 n};-\frac{1}{B}\right] \right. \right.\right.
\nonumber \\
   && \left. \left.\left. \times\left(\frac{B+t^{
\frac{6 n}{6 n-1}}}{B}\right)^n \left(t^{\frac{6 n}{6 n-1}}
   \left(B+t^{\frac{6 n}{6 n-1}}\right)\right)^{-n} \, _2F_1
\left[-n+1-\frac{1}{6
   n},n;-n+2-\frac{1}{6 n};-\frac{t^{\frac{6 n}{6 n-1}}}{B}
\right]\right)\right)\right)\,,\nonumber\label{dl}
\end{eqnarray}
where ${}_2F_1$ is an hypergeometric function. On inverting the relation $z(t)$ in Eq. (\ref{zt}), we obtain
\begin{eqnarray}
&&t(z)=2^{\frac{1}{6 n}-1} \left(\left((z+1) (B+1)^{-n}
\right)^{-1/n}\right. \\
&&\left.\times\sqrt{\left((z+1)
   (B+1)^{-n}\right)^{\frac{1}{n}} \left(B^2 \left((z+1)
   (B+1)^{-n}\right)^{\frac{1}{n}}+4\right)}-B\right)
^{1-\frac{1}{6 n}}\,,\nonumber\label{tz}
\end{eqnarray}
and we can construct $d_L(z)$ and evaluate the distance modulus according to Eq. (\ref{eq:modg}), so as to
perform our likelihood analysis, by maximizing the likelihood $\mathcal{L}=\exp{\left(  -\frac{1}
{2}\chi^{2}\right)}$ on a grid in the space of parameters\footnote{Let us note that in principle it could be
better to use the re--parametrization described above (i.e. the dependence of $B$ and $n$ on $H_0$ and
$\Omega_m$ provided by Eqs. (\ref{inv1}) and (\ref{inv2})) and to maximize the likelihood with respect to the
parameters $h$, $H_0$ and $\Omega_m$, since they have a brighter physical meaning and their space of parameters
is easily reconstructed as discussed above. However, the mathematical form of $B(H_0,\Omega_m)$ and
$n(H_0,\Omega_m)$, when inserted in the luminosity distance and in the correction term $\Delta M_{G}$, is such
that they make the numerical calculations too hard and time consuming. Thus, we prefer to work in the space of
parameters $h$, $B$ and $n$. Anyway, we will work in the space of parameters $h$, $H_0$ and $\Omega_m$ later,
when we implement the cosmographic approach. } $B$ and $n$. In order to constrain the parameters of our model
only, when we perform our statistical analysis with the Union2 compilation, we marginalize over  $h$, that is,
we maximize the likelihood $\mathcal{L}_{marg}=\int_{h_{min}}^{h_{max}}{d h \exp{\left(
-\frac{1}{2}\chi^{2}\right)  }}$, where $h_{min}$ and $h_{max}$ are fixed by using the latest WMAP7 results. We
obtain $\chi_{reduced}^2=0.97\,$ for 557 data-points and the regions of confidence at $3\, \sigma$ for $H_0$ and
$\Omega_m$ are $\left(0.92, 1.01\right)$ and $\left(0.24, 0.4\right)$, respectively. If we do not marginalize
over h, we obtain $h_{best}=0.70^{+0.02}_{-0.02}$, from which we can infer the following interval of confidence
at $3\, \sigma $: $\tau \in \left(12.8, 14.7\right)$ Gyr. In Fig. \ref{f3} we plot the Union2 data set with the
best fit modulus of distance, showing that they are, indeed, well-fitted by our model.

\section{Constraints from \textit{calibrated} Gamma Ray Bursts Hubble diagram}

Let us now assess another set of observations which we think essential to begin to understand how much our
theoretical model can be considered as a reliable one. As a matter of fact, it has been recently empirically
established that some of the directly observed parameters of Gamma Ray Bursts are connected with the isotropic
absolute luminosity $L_{iso}$, the collimation corrected energy $E_{\gamma}$, or the isotropic bolometric energy
$E_{\rm iso}$ of a GRB. Such observable properties of the GRBs include the peak energy, denoted by $E_{\rm
p,i}$, which is the photon energy at which the $\nu\,F_{\nu}$ spectrum is brightest; the jet opening angle,
denoted by $\theta_{\rm jet}$, which is the rest-frame time of the achromatic break in the light curve of an
afterglow; the time lag, denoted by $\tau_{lag}$, which measures the time offset between high and low energy GRB
photons arriving on Earth; and the variability, denoted by $V$, which is the measurement of the
\textit{spikiness} or \textit{smoothness} of the GRB light curve. In the literature, there is a wide variety of
choices for the definition of $V$ \cite{Schaefer} in which the observed $V$ value varies as the inverse of the
time stretching, so the corresponding measured value should be multiplied by a correcting factor $(1+z)$. An
additional luminosity indicator is the minimum rise time \cite{Schaefer} denoted by $\tau_{rt}$, and taken to be
the shortest time over which the light curve rises by half the peak flux of the pulse.

These quantities appear to correlate with the GRB isotropic luminosity, its total collimation-corrected or its
isotropic energy. This property cannot be measured directly but rather it can be obtained through the knowledge
of either the bolo-metric peak flux, denoted by $P_{bolo}$, or the bolo-metric fluence, denoted by $S_{bolo}$.
Therefore, the isotropic luminosity is given by
\begin{equation}
 L_{\rm iso} = 4\pi d^2_{L}(z)P_{\rm bolo} \,,\label{ldl}
\end{equation}
the total isotropic energy reads as
\begin{equation}
E_{\rm iso}=4\pi d^2_{L}(z)S_{\rm bolo}(1+z)^{-1}\,,
\end{equation}
and the total collimation-corrected energy is
\begin{equation}
E_{\gamma}=4\pi d^2_{L}(z) S_{\rm bolo}F_{\rm beam}(1+z)^{-1}\,, \label{egdl}
\end{equation}
where $F_{\rm beam}$ is the beaming factor. The correlation relations are power-law relations of either $ L_{\rm
iso}$ or $E_{\gamma}$ or $ E_{\rm iso}$ as a function of $\tau_{\rm lag}$, $V$,\, $E_{\rm peak}$,\, $\tau_{rt}$,
i.e.
\begin{eqnarray}\nonumber
&&E_{\rm iso} = b_{\rm iso,peak} E_{\rm peak}^{a_{\rm iso,peak}}\,,\nonumber\\
&&E_{\gamma} = b_{\gamma,peak}E_{\rm peak}^{a_{\gamma,peak}}\,,\nonumber\\
&&L = b_{peak}E_{\rm peak}^{a_{peak}}\,.\nonumber\\
\end{eqnarray}
Therefore, $L_{iso}$, $E_{\gamma}$ and $ E_{\rm iso}$ depend not only on the GRB observables $P_{\rm bolo}$ or
$S_{\rm bolo}$, but also on the cosmological parameters, through the luminosity distance $d_L(z)$. As a
consequence, there is a fierce problem to overcome, since it is not immediately possible to calibrate such GRBs
empirical laws, and to build up a new GRBs Hubble diagram, without assuming any a priori cosmological model
(which is known as the circularity problem).

\begin{figure}
\includegraphics[width=8 cm, height=6 cm]{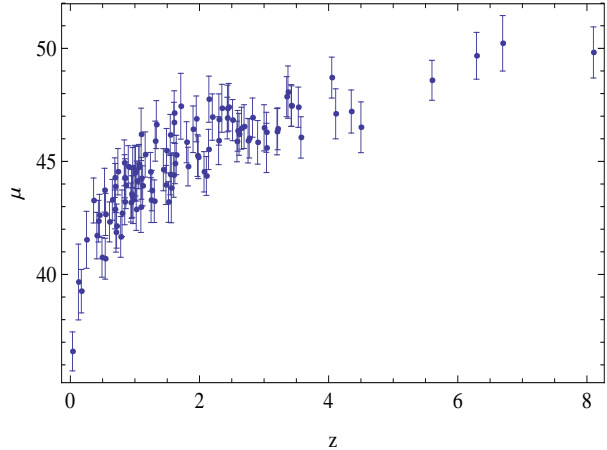}
\caption{Distance modulus $\mu(z)$ for the \textit{calibrated}  GRBs Hubble diagram made up by fitting the Amati
correlation.} \label{figrbs}
\end{figure}

\begin{figure}
\includegraphics[width=8 cm, height=6 cm]{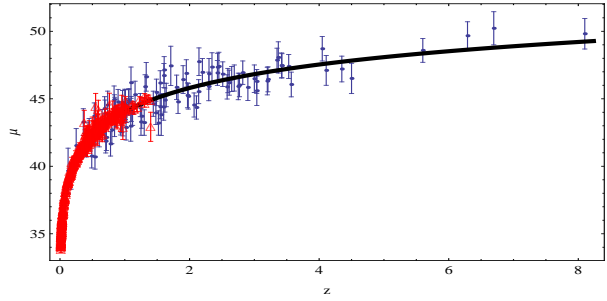}
\caption{The \textit{calibrated} GRBs Hubble diagram with overplotted the distance modulus predicted by the
fiducial model (solid line). The full circles correspond to the GRBS data set, while the empty red triangles
correspond to the Union2 SneIa data points.} \label{hdcal}
\end{figure}

In Refs. \cite{MEC10} and \cite{ME} we have applied a local regression technique to estimate, in a model
independent way, the distance modulus from the recently updated Union SNeIa sample, containing $557$ SNeIa
spanning the redshift range of $0.015 \le z \le 1.55$. The derived calibration parameters have been used to
construct an updated GRBs Hubble diagram. In particular, by using such a technique, we have fitted the so-called
\textit{Amati relation} (see Ref. \cite{MEC10} for details) and constructed an updated Gamma Ray Bursts Hubble
diagram, which we call the \textit{calibrated} GRBs HD, consisting of a sample of $109$ objects, shown in Fig.
\ref{figrbs}. Their redshift distribution covers a broad range of $z$, from $0.033$ to $8.23$, thus extending
far beyond that of SNeIa ($z <\sim1.7$), and including GRB $092304$, the new high-z record holder of Gamma Ray
Bursts.
Here we want to use such \textit{calibrated} GRBs HD  to test if our cosmological model is able to describe the
background expansion up to  redshifts $z\sim 8$. In our Bayesian approach to model testing, we explore the
parameter space through the likelihood function
\begin{equation}
{\cal{L}}_{GRB}({\bf p}) \propto \exp{[-\chi_{GRB}^2({\bf p})/2]}\,, \label{eq:likegrbsimple}
\end{equation}
with
\begin{equation}
\chi_{GRB}^2({\bf p}) = \sum_{i = 1}^{{\cal{N}}_{GRB}}{\left [ \frac{\mu_{obs}(z_i) -
\mu_{th}(z_i)}{\sqrt{\sigma_i^2 + \sigma_{GRB}^2}} \right ]^2}\,, \label{eq: defchigrb}
\end{equation}
where $\sigma_{GRB}$ takes into account the intrinsic scatter inherited from the scatter of GRBs around the
Amati correlation (see Ref. \cite{MEC10} and references therein), {\bf p} denotes the set of model parameters
($B$ and $n$ and $h$ in our case), and the distance modulus $\mu(z)$ is provided by Eq. (\ref{eq:mMr}). The
inferred confidence intervals (at $3\,\sigma $) for $H_{0}$, $\Omega_m$ and $h$, are $H_0 \in \left(0.96,
1.1\right)$, $\Omega_m \in \left(0.26, 0.39\right)$, and $h \in \left(0.65,0.74\right)$. We obtain
$\chi_{red}^2=0.97$ for $109$ data points. In Fig. \ref{hdcal} we show the GRBs Hubble diagram with overplotted
the distance modulus predicted by the fiducial model. It turns out that our cosmological model is fully
compatible with this recently compiled GRBs HD.

\section{Constraints from Chandra X-ray observations of
large relaxed galaxy clusters}

The matter content of the largest clusters of galaxies is expected to provide an almost fair sample of the
matter content of the Universe (see, for instance, Refs. \cite{white93}and \cite{allen08}). The ratio of
baryonic-to-total mass in clusters should, therefore, closely match the ratio of the cosmological parameters
$\Omega_{\rm b}/\Omega_{\rm m}$. The baryonic mass content of clusters is dominated by the X-ray emitting gas,
the mass of which exceeds the mass of optically luminous material by a factor $\sim 6$, with other sources of
baryonic matter being negligible. The combination of robust measurements of the baryonic mass fraction in
clusters from X-ray observations together with a determination of $\Omega_{\rm b}$ from other measuremets (as
for instance cosmic microwave background (CMB) data or big-bang nucleosynthesis calculations) and a constraint
on the Hubble constant, can therefore be used to measure $\Omega_{\rm m}$ and constrain the parameters which
characterize any cosmological model. This constraint originates from the dependence of the $f_{\rm gas}$
measurements, which derive from the observed X-ray gas temperature and density profiles, on the assumed
distances to the clusters, $f_{\rm gas} \propto d^{1.5}$.

To understand the origin of the $f_{\rm gas} \propto d^{1.5}$ dependence, consider a spherical region of
observed angular radius $\theta$,  within which the mean gas mass fraction is measured. The physical size, $R$, is
related to the angle $\theta$ as $R = \theta d_{\rm A}$.  The X-ray luminosity emitted from within this region,
$L_{\rm X}$, is related to the detected flux, $F_{\rm X}$, as $L_{\rm X} = 4\pi d_{\rm L}^2 F_{\rm X}$, where
$d_{\rm L}$ is the luminosity distance and $d_{\rm A} =d_{\rm L}/(1+z)^2$ is the angular diameter distance.
Since the X-ray emission is primarily due to collisional processes (bremsstrahlung and line emission) and is
optically thin, we may also write $L_{\rm X} \propto n^2 V$, where $n$ is the mean number density of colliding
gas particles and $V$ is the volume of the emitting region, with $V=4\pi (\theta d_{\rm A})^3/3$. On considering
the cosmological distance dependences, we see that $n \propto d_{\rm L}/d_{\rm A}^{1.5}$, and that the observed
gas mass within the measurement radius $M_{\rm gas} \propto nV \propto d_{\rm L}d_{\rm A}^{1.5}$. The total
mass,  $M_{\rm tot}$, determined from the X-ray data under the assumption of hydrostatic equilibrium, is such
that $M_{\rm tot} \propto d_{\rm A}$. Thus, the X-ray gas mass fraction measured within angle $\theta$ is
$f_{\rm gas}=M_{\rm gas}/M_{\rm tot} \propto d_{\rm L}d_{\rm A}^{0.5}$. The expectation from non-radiative
hydrodynamical simulations is that for the largest ($kT \succeq 5$\,keV), dynamically relaxed clusters and for
measurement radii beyond the innermost core ($r \succeq r_{2500}$), $f_{\rm gas}$ should be approximately
constant with redshift, the virial radius $r_{2500}$  being defined as the radius of a sphere such that the mean
density contained within it is $\Delta= 2500$ times the critical density at the halo redshift.

It is worth noting that even if the virial radius $r_{2500}$ depends on the fiducial cosmological model, in
order to determine constraints on cosmological parameters it is not necessary to generate $f_{\rm gas}$ datasets
for every cosmology of interest and compare them with the expected behaviour. Indeed, it is possible to fit a
single \textit{fiducial} $f_{\rm gas}$ dataset with a model that accounts for the expected apparent variation in
$f_{\rm gas}(z)$ as the underlying cosmology is varied. Let us choose the $\Lambda$CDM reference cosmology.
Following Ref. \cite{allen08}, the model fitted to the reference $\Lambda$CDM data is
\begin{equation}
f_{\rm gas}^{\rm \Lambda CDM}(z) = \frac{ K \mathcal{A} \gamma b(z)} {1+s(z) } \left( \frac{\Omega_{\rm
b}}{\Omega_{\rm m}} \right) \left[ \frac{d_{\rm A}^{\rm \Lambda CDM}(z)}{d_{\rm A}(z)} \right]^{1.5}\,,
\label{eq:fgas}
\end{equation}
where $d_{\rm A}(z)$ and $d_{\rm A}^{\rm \Lambda CDM}(z)$ are the angular diameter distances to the clusters in
our test and reference cosmological models, respectively.

In order to construct the angular diameter distance for our cosmological model, we use the relation between the
angular diameter distance $d_{\rm A}$ and the luminosity distance $d_L$
\begin{equation}\label{lum2}
d_{\rm L}=\left(1+z\right)^2 d_{\rm A}\,,
\end{equation}
$d_{\rm L}$ being given in Eq. (\ref{dl}). In Eq. (\ref{eq:fgas}) $\mathcal{A}$ takes into account the change in
angle subtended by $r_{2500}$ as the underlying cosmology is varied, and can be evaluated as
\begin{equation}
\mathcal{A}= \left( \frac{ \theta_{2500}^{\rm \Lambda CDM}} {\theta_{2500}} \right)^\eta \approx \left( \frac{
H(z) d_{\rm A}(z)~~~~~~~} { \left[ H(z) d_{\rm A}(z)\right] ^{\rm \Lambda CDM}} \right)^\eta\,.
\label{eqn:angcorrection}
\end{equation}
We take the value of the slope, $\eta$, of the $fgas(r/r_{2500})$ data in the region of $r_{2500}$ (as measured
for the reference $\Lambda$CDM cosmology), indicated in Ref. \cite{allen08}, that is $\eta=0.214\pm0.022$. The
parameter $\gamma$ in Eq. (\ref{eq:fgas}) models non-thermal pressure support in the clusters. On relying upon
hydrodynamical simulations, we take $1.0<\gamma<1.2$. The parameter $s(z)=s_0(1+\alpha_{\rm s}z)$ in Eq.
(\ref{eq:fgas}) models instead the contribution to the baryonic mass given by stars. The factor $b(z)=b_0 (1+
\alpha_{\rm b}z)$ is the `depletion' or `bias' factor and describes, in a completely \textit{empirical} way, the
ratio by which the baryon fraction measured at $r_{2500}$ is depleted with respect to the universal mean.
According to Ref. \cite{allen08} we choose $s_0=(0.16 \pm 0.05)h_{70}^{0.5}$, $-0.2 <\alpha_{\rm s}<0.2$,
$0.65<b_0<1.0$ and $-0.1<\alpha_b<0.1$ (which corresponds to a moderate, systematic evolution in $b(z)$). The
factor $K$ in Eq. (\ref{eq:fgas}) is a calibration constant fixed to $K=1.0\pm 0.1$.

\begin{figure}
\includegraphics[width=8 cm, height=5 cm]{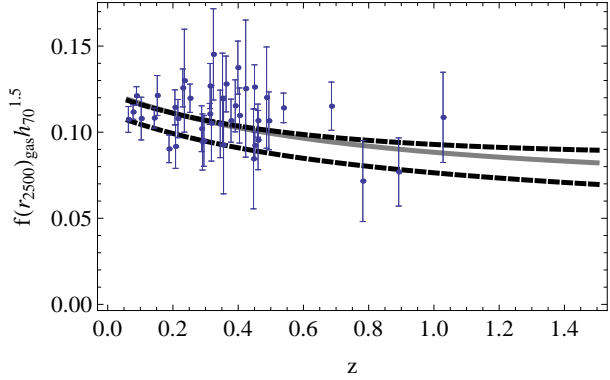}
\caption{The variation of the X-ray gas mass fraction measured within $r_{2500}$ as a function of redshift for
our model. The dashed lines, which outline the \textit{maximum likelihood} region, correspond to the
\textit{extreme} values for the parameters $\alpha_{\rm s}$ and $\alpha_{\rm b}$, while the gray solid line
corresponds to $\alpha_{\rm s}=\alpha_{\rm b}=0$.   } \label{fgasgas}
\end{figure}

As above, we perform a Bayesian analysis, maximizing our likelihood $\mathcal{L}=\exp{\left( -\frac{1}
{2}\chi^{2}\right)  }$ on a grid in the space of parameters $B$ and $n$, and varying all the astrophysical
parameters appearing in Eq. (\ref{eq:fgas}). Moreover, as in Ref. \cite{allen08}, we use the standard priors
with $\Omega_{\rm b} h^2=0.0214 \pm 0.0020$ and $h=0.72 \pm 0.08$. We obtain, in such a way, a sort of
\textit{maximum likelihood region}, where $f( r_{2500})_{\mathbf{ gas}} h^{1.5}$ can vary, as shown in Fig.
\ref{fgasgas}.
The inferred region of confidence (at $3\, \sigma $) for $H_0$ and $\Omega_{\rm m}$ are $\left(0.85,
1.05\right)$ and $\left(0.26, 0.51\right)$, respectively.

\section{Cosmography}

\subsection{General approach}

Over the last years the cosmographic approach to cosmology gained increasing interest for catching as much
information as possible directly from observations, retaining the minimal priors of isotropy and spatial homogeneity and
leaving aside any other assumptions. Actually, the only ingredient taken into account \textit{a priori} in this
approach  is the FLRW line element obtained from kinematical requirements
\begin{equation}
ds^2=-c^2dt^2+a^2(t)\left[\frac{dr^2}{1-kr^2}+r^2d\Omega^2\right]\,.
\end{equation}
By using this metric, it is possible to express the luminosity distance $d_L$ as a power series in the redshift
parameter $z$, the coefficients of the expansion being functions of the scale factor $a(t)$ and its higher-order
derivatives. Such an expansion leads to a distance\,-\,redshift relation which only relies on the assumption of
the Friedmann--Lemaitre--Robertson--Walker metric, thus being fully model independent since it does not depend
on the particular form of the solution of cosmic equations. For this purpose, it is convenient to introduce the
following parameters:
\begin{eqnarray}
H &=& \frac{1}{a} \frac{da}{dt}\,, \\
q &=& - \frac{1}{aH^{2}} \frac{d^2a}{dt^2}\,,\\
j &=& \frac{1}{aH^{3}} \frac{d^3a}{dt^3}\,, \\
s &=& \frac{1}{aH^{4}} \frac{d^4a}{dt^4}\,. \\ \label{eq: cosmopar}
\end{eqnarray}
These parameters are usually referred to as the Hubble, deceleration, jerk\footnote{The use of the jerk
parameter to discriminate between different models was also proposed in the context of the {\it statefinder}
parametrization \cite{sahni03}.}, and snap parameters, respectively.

Their present day values (which, as usual, we will denote with a subscript $0$) can be used to characterize the
evolutionary status of the Universe. For example, $q_0 < 0$ denotes an accelerated expansion, while a change of
sign of $j$ (in an expanding universe) signals that the acceleration starts increasing or decreasing. Most
importantly, the parameters $\{q_0,j_0,s_0\}$ can be used to evaluate different distances in the universe. This
can be achieved by inverting the scale factor series expansion (approximated to the fourth order in $t-t_0$) of
a FLRW metric in terms of time, given in the following equation:
\begin{equation}
\frac{a(t)}{a(t_0)}=1+H_0(t-t_0)-\frac{q_0}{2}H_0^2(t-t_0)^{2} +
\frac{j_0}{3!}H_0^3(t-t_0)^3+\frac{s_0}{4!}H_0^4(t-t_0)^{4}\,. \label{expansionat}
\end{equation}
Therefore, one can obtain a series expansion of a distance, $D(t_1,t_0)$, travelled by a given photon that was
emitted at $t_1$ and detected at the current epoch $t_0$, in terms of the scale factor or redshift, while the
coefficients of the expansion are defined through the cosmographic parameters. Such a distance can be related to
several physical quantities, for example the luminosity distance, the angular diameter distance and more. These
quantities can be constrained observationally through, for example, SNeIa, gravitational lenses, and possibly,
GRB data. It is worth noticing, in this respect, that since the cosmographic approach is based on a Taylor
expansion of the scale factor, or redshift, for data of GRB at high redshift (above $z=1$), it is better to use
the variable  $y=z/(1+z)$, introduced in Ref. \cite{polarsky}, in such a way that $z\in(0,\infty)$ is mapped
into $y\in(0,1)$, obtaining
\begin{eqnarray}
d_L(y)=&&\frac{c}{H_0}\left\{y-\frac{1}{2}(q_0-3)y^2 +\frac{1}{6}\left[12-5q_0+3q^2_0-j_0\right]y^3
+\frac{1}{24}\left[60-7j_0\right.\right.\nonumber
  \\ &&\left.\left.-10-32q_0+10q_0j_0+6q_0+21q^2_0-15q^3_0
+s_0\right]y^4+\mathcal{O}(y^5)\right\}~\,.
\end{eqnarray}

\subsection{Application of cosmography to our model}

In this subsection, we will relate the parameters characterizing the model introduced in Sec. II to the
cosmographic parameters $\{q_0,j_0,s_0\}$.  By expanding our approximate luminosity distance up to the fourth
order in the $y$-parameter, and comparing such an expansion with the \textit{standard} expansion to the fourth
order, we get the \textit{map} which relates the parameters $B$ and $n$ of our model to the cosmographic
parameters $q_0$, $j_0$, $s_0$. Actually, we find
\begin{eqnarray}
q_0&=&\frac{6 B (B+2) n-B (B+3)+12 n-2}{6 (B+2)^2 n^2}-1\,,\\
\nonumber\\
j_0&=&18 \left((B+2)^3 n^4-54 (B+2) (B (B+2)+2) n^3+9 (B+2)
(B (5 B+7)+6) n^2\right.\nonumber\\
&-& \left. 3
   (B+1) (B (4 B+9)+8) n+(B+1)^2 (B+2)\right) \frac{1}{18 (B+2)^3 n^4}\,,\\
   \nonumber\\
s_0 &=& \left[B^4 (2 (n-3) n+1) (3 (n-2) n+1) (6 (n-1) n+1)
+ \right.\nonumber\\
&+& \left.B^3 (n (6 n (6 n (n
   (4 n (2 n-9)+59)-43)+85)-79)+5)+\right.\\
   &+&\left.B^2 (n (n (24 n (9 n (2 n (2
   n-7)+19)-112)+851)-140)+9)+\right.\nonumber \\
   &+&\left. B (n (72 n (n (2 n (8 (n-3)
   n+27)-31)+10)-115)+7) + \right.\nonumber\\
   &+&\left. 2 (6 (n-1) n+1) \left(4 n^2-6 n+1\right)
   (6 n (2 n-1)+1)\right]\frac{1}{36 (B+2)^4 n^6}\,.\nonumber
\end{eqnarray}

In order to constrain the model, we need to constrain observationally the cosmographic parameters by using
appropriate distance indicators. Moreover, we must take care that the expansion of the distance related
quantities in terms of $(q_0, j_0, s_0)$ closely follows the exact expressions over the range probed by the data
used. Taking SNeIa and a fiducial $\Lambda$CDM model as a test case, one has to check that the approximated
luminosity distance deviates from the $\Lambda$CDM one by less than the measurement uncertainties up to $z
\simeq  1.5$, to avoid introducing any systematic bias. Since we are interested in constraining the cosmographic
parameters, we will expand the luminosity distance $D_L$ up to the fifth order in $z$ which indeed allows us to
track the $\Lambda$CDM expression with an error less than $1\%$ over the full redshift range. To constrain the
parameters $(h, q_0, j_0, s_0)$, we define the likelihood ${\cal{L}}({\bf p})$ as\,
\begin{eqnarray}
{\cal{L}}({\bf p}) & \propto & \frac{\exp{(-\chi^2_{SneIa/GRB}/2)}}{(2 \pi)^{\frac{{\cal{N}}_{SneIa/GRB}}{2}}
|{\bf C}_{SneIa/GRB}|^{1/2}} \nonumber \\
~ & \times  & \frac{1}{\sqrt{2 \pi \sigma_h^2}} \exp{\left [ - \frac{1}{2} \left ( \frac{h - h_{obs}}{\sigma_h}
\right )^2\right ]} \nonumber \\
~ & \times & \frac{\exp{(-\chi^2_{BAO}/2})}{(2 \pi)^{{\cal{N}}_{BAO}/2} |{\bf C}_{BAO}|^{1/2}} \nonumber \\
~ & \times & \frac{1}{\sqrt{2 \pi \sigma_{{\cal{R}}}^2}} \exp{\left [ - \frac{1}{2} \left ( \frac{{\cal{R}} -
{\cal{R}}_{obs}}{\sigma_{{\cal{R}}}} \right )^2 \right ]} \nonumber \\
~ & \times & \frac{\exp{(-\chi^2_{H}/2})}{(2 \pi)^{{\cal{N}}_{H}/2} |{\bf C}_{H}|^{1/2}} \,,
\label{defchiall}
\end{eqnarray}
where
\begin{eqnarray}
\chi_{SneIa/GRB}^2({\bf p}) & = & \sum_{i = 1}^{{\cal{N}}_{SneIa/GRB}}{\left [ \frac{\mu_{obs}(z_i) - \mu_{th}(z_i,
{\bf p})}{\sigma_i} \right ]^2}. \nonumber \\
\label{defchiSneIa}
\end{eqnarray}
As observational dataset we actually use the currently available observational SNIa and GRB Hubble Diagrams, and
we set Gaussian priors on the distance from Baryon Acoustic Oscillations (BAO), and the Hubble constant $h$.
Such priors have been included in order to help break the degeneracies among the parameters of the cosmographic
series expansion. In Eq. (\ref{defchiall}) ${\bf C}_{SneIa/GRB}$ is the SneIa/GRBs diagonal covariance matrix
and $(h_{obs}, \sigma_h) = (0.742, 0.036)$. The third term takes into account the constraints on $d_z =
r_s(z_d)/D_V(z)$ with $r_s(z_d)$ the comoving sound horizon at the drag redshift $z_d$ (which we fix to be
$r_s(z_d) = 152.6 \ {\rm Mpc}$ from WMAP7) and the volume distance is defined as in Ref. \cite{Eis05}:
\begin{equation}
D_V(z) = \left \{ \frac{c z}{H(z)} \left [ \frac{D_L(z)}{1 + z} \right ]^2 \right \}^{1/3} \ .
\label{eq: defdv}
\end{equation}
The values of $d_z$ at $z = 0.20$ and $z = 0.35$ have been estimated in Ref. \cite{percival10} using the SDSS
DR7 galaxy sample so that we define $\chi^2_{BAO} = {\bf D}^T {\bf C}_{BAO}^{-1} {\bf C}$ with ${\bf D}^T =
(d_{0.2}^{obs} - d_{0.2}^{th}, d_{0.35}^{obs} - d_{0.35}^{th})$ and ${\bf C}_{BAO}$ the BAO covariance matrix.
The next term refers to the shift parameter \cite{B97} \cite{EB99}:
\begin{equation}
{\cal{R}} = \sqrt{\Omega_M} \int_{0}^{z_{\star}}{\frac{dz'}{E(z')}}
\label{eq: defshiftpar}
\end{equation}
with $z_\star = 1090.10$ the redshift of the last scattering surface. We follow again WMAP7 by setting
$({\cal{R}}_{obs}, \sigma_{{\cal{R}}}) = (1.725, 0.019)$. Thus we obtain the following confidence region (at
$3\sigma$) for the cosmographic parameters: $\left(
\begin{array}{ccc}
 q_0& -0.76 & -0.25 \\
 j_0&0.28& 0.33 \\
 s_0& -2.15 & -1.18\,
\end{array}
\right)$\,
confirming an accelerated stage of the universe.

\begin{figure}
\includegraphics[width=8 cm, height=5 cm]{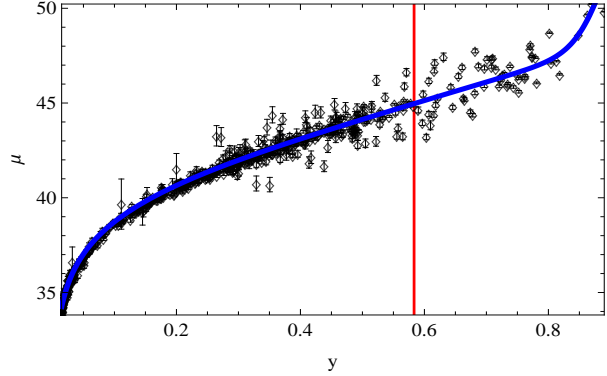}
\caption{Distance modulus for the best-fit values of our cosmography (solid blue line) performed with the
\textit{cosmographic dataset} (empty black diamond). The red vertical line indicates the \textit{extreme} value
of the parameter $y=z/(1+z)$ where we have SNeIa data.} \label{cosmogrball}
\end{figure}

In addition, we can investigate the possibility to use high redshift GRBs to determine parameters of our
cosmography. As a matter of fact, we use a whole dataset containing both the SNIa Union2 dataset and the
\textit{calibrated} GRBs HD, which we call the \textit{cosmographic dataset}. We obtain the following confidence
region (at $3\sigma$) for the cosmographic parameters: $\left(
\begin{array}{ccc}
 q_0& -0.6 & -0.2 \\
 j_0&0.01& 0.21 \\
 s_0& -2.14 & 0.86\,
\end{array}
\right)$. In Fig. \ref{cosmogrball} we actually show the \textit{cosmographic distance modulus} together with
the \textit{cosmographic dataset}.

As a final step, we can use the re--parametrization of our model in terms of $H_0$ and $\Omega_m$ exhibited in
the subsection (\ref{rep}) to construct $d_{L}^{\rm{cosmographic}}(y,H_0,\Omega_m, h)$, and perform the
cosmographic analysis using both Supernovae and Gamma Ray Bursts data (the so-called \textit{cosmographic
dataset}), which allow us to obtain constraints (even if not yet stringent) on the parameters of cosmography.
Actually, it turns out that
\begin{equation}
H_0= 0.98^{+0.04}_{-0.05}\,,\,\,\,\,\,\,\, \Omega_m=0.2\pm 0.05\,,
\end{equation}
and then, from Eqs. (\ref{inv2}), we obtain the following confidence region
(at $3\sigma$) for the cosmographic parameters:
$\left(
\begin{array}{ccc}
 q_0& -0.77 & -0.26 \\
 j_0&0.11& 0.39 \\
 s_0& -2.2 & -0.64\,
\end{array}
\right)$. In Fig \ref{cosmogrball} we plot the distance modulus for the best-fit values of our cosmography
(solid blue line) performed with the \textit{cosmographic dataset} (empty black diamond). The red vertical line
indicates the \textit{extreme} value of the parameter $y=z/(1+z)$ where we have SNeIa data.

\section{A more realistic description}

The model of the universe adopted so far is described by an exact solution of the dynamical equations, whose arbitrary
parameters are determined by specifying the initial conditions. In this section we now consider more realistically the inclusion of radiation, too, into such a model. In this case the dynamical equations, as far as we know, do not have analytical solutions, and therefore we will rely on numerical
computations.

The dynamical field equations become
\begin{equation}
\frac{\ddot{a}}{a} + \frac{\dot{a}^2}{2a^2}  - \frac{\Lambda}{2} - \frac{\dot{a}\dot{G}}{a G} + \frac{\mu
\dot{G}^2}{4G^2} + 4\pi G(\gamma_{rad}-1)D_{rad}a^{-3\gamma_{rad}}=0\,, \label{eqrad1}
\end{equation}
\begin{equation}
\mu \ddot{G} - \frac{3}{2}\mu \frac{{\dot{G}}^2}{G} + 3\mu \frac{\dot{a}}{a}\dot{G} + \frac{G}{2} \left(
-6\frac{\dot{a}^2}{a^2}-2\Lambda + 2G\frac{d\Lambda}{d G} \right) = 0\,. \label{eqrad2}
\end{equation}
We have also to consider the Hamiltonian constraint \cite{Bona04}
\begin{equation}
\frac{\dot{a}^2}{a^2} - \frac{\Lambda}{3} - \frac{\mu}{6}\frac{\dot{G}^2}{G^2} -\frac{8\pi
G}{3}(D_{m}a^{-3\gamma_{m}}+D_{rad}a^{-3\gamma_{rad}})= 0\,.\label{eqrad3}
\end{equation}
It turns out that these equations assume a simpler form when, instead of $t$ as an independent variable, one
uses $a(t)$ - the scale factor. We introduce a new independent variable by $u \equiv \log(1+z)=-\log({a(t)\over
a_{0}})$, where $a_{0}$ is the present value of the scale factor (fixed at $a_0=1$) and $z$ is the redshift. The
equations can now be written in the form
\begin{equation}
H^2(u)\left(1-\frac{2}{3}\mu \frac{H'(u)}{H(u)}\right) = \frac{\Lambda(u)}{3} + \frac{2}{3}\mu H^2(u)
\left(\frac{G'(u)}{G(u)}\right)^2+ \frac{8 \pi
}{3}G(u)D_{rad}a^{-3\left(\gamma_{rad}-1\right)}\,,\label{eqrad1b}
\end{equation}
\begin{equation}
G''(u) = G'(u)\left( 3+\frac{H'(u)}{H(u)}+\frac{3}{2}\frac{G'(u)}{G(u)} \right)+ \frac{G(u)}{\mu}\left(
-3+\frac{3n\Lambda(u)}{(1-3 n)H^2(u)} \right)\,, \label{eqrad2b}
\end{equation}
\begin{equation}
H^2(u) = \frac{\frac{8 \pi G(u)}{3}\left( D_{rad}a^{-3\gamma_{rad}}+D_{m}a^{-3\gamma_{m}} \right)+
\frac{\Lambda}{3}}{1+\frac{\mu}{6}\frac{G'(u)}{G(u)}^2}\,, \label{eqrad3b}
\end{equation}
where $'=\frac{d}{du}$, $\gamma_{rad}=\frac{4}{3}$, and $\gamma_{m}=1$.

\begin{figure}
\includegraphics[width=7 cm]{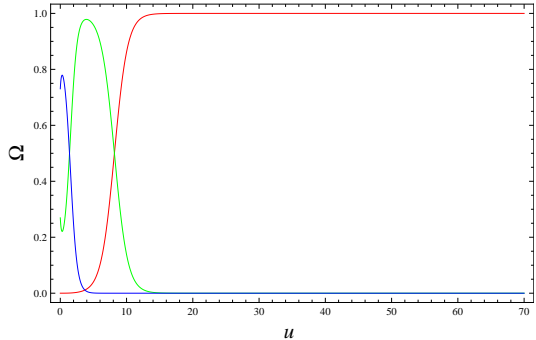}
\caption{$\Omega$ parameters as functions of $u$
        in the universe filled in with matter, radiation and the $\Lambda$-term.
        $\Omega_{\Lambda}$ is marked in blue, $\Omega_{r}$ in red and $\Omega_{m}$ in green.} \label{omegarad}
\end{figure}

\begin{figure}
\includegraphics[width=6 cm, height=6 cm]{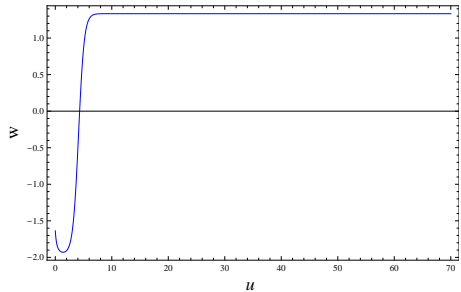}
\caption{The effective equation of state $w$ as a function of $u$.} \label{wlg}
\end{figure}

Moreover, we have to remember that above, in Eq. (\ref{eqrad2b}), we explicitly used the  dependence of
$\Lambda$ on $G$, as given by applying the Noether Symmetry Approach to the case of a matter--dominated universe
(without any radiation content)\footnote{We actually remember that in the presence of radiation it is not
possible to solve analytically the equations by defining the existence of the Noether symmetry vector field. }:
$\Lambda(u)= WG^{\frac{1}{1-3n}}(u)$. Let us note that Eqs. (\ref{eqrad1b}) and (\ref{eqrad3b}) can be rewritten
in a \textit{Friedmann-like} form
\begin{equation}
H^2(u)\left( 1-\frac{2}{3}\frac{H'(u)}{H(u)} \right) = -\frac{4\pi G }{3}\left(p_{\Lambda, G} +p_{rad}\right)\,,
\label{eqrad1c}
\end{equation}
\begin{equation}
H^2(u)= \frac{8 \pi G(u)}{3}\left(\rho_{\Lambda,G}+\rho_{rad}+\rho_m\right)\,,\label{eqrad3c}
\end{equation}
where we define
\begin{equation}
\rho_{\Lambda,G}\equiv\displaystyle\frac{\Lambda}{8 \pi G(u)}+\frac{H(u)^2 \left( 3-\frac{\mu}{6} \right)
G'(u)^2}{4\pi  G^3}\,,\label{density}
\end{equation}
\begin{equation}
p_{\Lambda,G}\equiv\displaystyle\frac{H^2
 \left( 3-\frac{2\mu}{3} \right) u^2 \left( G' \right)^2}{4 \pi G^3}-\frac{\Lambda}{4\pi G}\,, \label{pressure}
\end{equation}
and then construct $w=\displaystyle \frac{p_{\Lambda,G}}{\rho_{\Lambda,G}}$.
Substituting Eq. (\ref{eqrad3b}) in Eq. (\ref{eqrad2b}), we can numerically solve the system of differential
equations characterizing cosmology, by specifying the initial condition at $u=70$, for example, and assuming
that $G(70)$, and ${G'(70)}$ have the same values as in the case without radiation. As can be seen in Fig.
\ref{omegarad}, the presence of radiation affects the evolution of the $\Omega$ parameters. As
expected, at the initial time, again when $u=30$, radiation dominates the expansion rate of the universe, with
the $\Lambda$-term and matter being subdominant, at a redshift $z$ of about $5000$; the energy densities of
matter and radiation become equal, and, for a relatively short period, the universe becomes matter dominated
until, at a redshift of about $1$, the $\Lambda$-term starts dominating the expansion rate of the universe.
In Fig. \ref{wlg} we plot the behaviour of $w$ as a function of $u$: we can therefore see that it behaves like
stiff matter ($w=\frac{4}{3}$) up to $u\simeq 10$, when there is the beginning of a transition towards a
\textit{superquintessence} behavior with $w< -1$.

It is worth noting that, since the initial conditions needed to numerically integrate the Eqs. (\ref{eqrad1c}),
(\ref{eqrad3c}) and (\ref{density}), have been set with the only constraint of obtaining a well--behaved
evolution for the $\Omega$ parameters, the values $w(0)$, $\Omega_{\Lambda}(0)$, $\Omega_{r}(0)$, and
$\Omega_{m}(0)$, cannot be directly compared with the ones indicated in the literature (see for instance Ref.
\cite{amanullah}), because they do not result from a fitting procedure on observational datasets. Moreover, as
far as the equation of state of dark energy is concerned, the best--fit values also depend on the mathematical
law assumed for it. For instance, if we adopt models of dark energy as potential energy of some, as yet
undiscovered, scalar field, we cannot obtain \textit{superquintessential} values $w<-1$, unless we consider
phantom fields or non--minimal coupling terms. On the other hand, such values are fully acceptable when we look
at the confidence regions obtained by fitting the data with a parametrized form of the equation of state;
however, this is not the same as saying that the scalar field is ruled out by the statistics.

\section{Conclusions}

We have shown that one can build a matter-dominated cosmological model with variable Newton parameter and
variable cosmological term which is compatible with the more updated observations of type Ia supernovae, the
gamma ray bursts Hubble diagram, and the gas mass fraction in X-ray luminous galaxy clusters. Moreover, we have
applied to such a cosmological model a cosmographic approach, which can help in selecting realistic models
without a priori choices that can be questionable. In performing our cosmographic analysis we set Gaussian
priors on the distance from Baryon Acoustic Oscillations (BAO), and the Hubble constant $h$. Such priors have
been included in order to help break the degeneracies among the parameters of the cosmographic series expansion.
A more realistic approach, considering the inclusion of a radiation component, seems also possible, even if it
has to be worked out only numerically.

Some questions remain, however, unsolved; for example, it is not clear enough what can be definitely said about
the effects of  $G$, which might point out to the violation of the strong equivalence principle, but not
necessarily of the Einstein equivalence principle. Here we have simply discussed the observational bounds on the
possible variations of the gravitational constant in the early universe, considering the best limit (at $3
\sigma$), $\delta G = 0.09 ^{+0.22}_{-0.19}$, obtained in the  literature. It turned out that our model can
satisfy such a best fit limit, provided that the $n$ parameter is appropriately chosen (the role of $B$ is only
marginal with respect to this strong test) to be compatible with the basic cosmological observations. We
postpone to a forthcoming paper the detailed analysis of the the dependence of the various elemental abundances
on the time variation of $G$ for our model. We expect that the BBN constraints are the main tool to restrict the
allowed region of the model parameters, so representing a necessary step in the feasibility study of the model
itself.

The concrete possibility to generate acceptably a previous radiation--dominated regime in the framework of
Renormalization--Group inspired cosmology has still to be proved. Unfortunately, the procedure of the Noether
Symmetry Approach does not work with a Lagrangian where the matter term is $L_m \equiv D(1 - a^{-1})$ (being
$\gamma = 1$ for dust and $\gamma = 4/3$ for radiation). However, as we have seen, a numerical integration of
the equations derived by using such an $L_m$, implies that, after inflation, such equations do give rise to a
period of radiation dominance followed by one of matter dominance. Later, a period of acceleration only
depending on variability of $\Lambda$ becomes the dominant one.

As far as we can see, we think that the work here presented may be considered as another important step towards
a more accurate confrontation of variable--G cosmologies with modern observations. Even if, as said in the
Introduction, such a comparison has begun earlier with many of the papers cited in the references, they all
indeed only contribute to show a first feasibility of some of such cosmological models, not only from a
theoretical but also from an observational point of view. Much work still remains to be done but the years to
come will hopefully help in further restricting the family of these and other possibly viable cosmological
models, mainly from the observational point of view. Actually, in a forthcoming paper we are going to
investigate the behavior of density  perturbations during the matter dominated stage.

\section*{Acknowledgments}
G. Esposito and C. Rubano are grateful to the Dipartimento di Scienze Fisiche of Federico II University, Naples,
for hospitality and support.

\end{document}